# Optimization and Testing of a novel Photoacoustic Spectromicroscopy Cell in the Mid-IR Spectrum

*Kirk Michaelian,[1] Chris Kelley,[2] Tor Pedersen,[3] Mark Frogley,[2] Timothy May,[3] Luca Quaroni,[3,4,5*] Gianfelice Cinque.[2]*

[1] CanmetENERGY, *Devon, Alberta T9G 1A8, Canada.*
[2]*Diamond Light Source, Diamond Light Source Ltd, Diamond House, Harwell Science and Innovation Campus, Didcot, Oxfordshire, OX11 0DE, United Kingdom*
[3]*Canadian Light Source, Canadian Light Source Inc. 44 Innovation Boulevard Saskatoon, S7N 2V3, SK, Canada.*
[4]*Institute of Nuclear Physics, Polish Academy of Sciences, ul. Radzikowskiego 152, 31-342 Kraków, Poland*
[5]*Jagiellonian University, Faculty of Chemistry, ul. Ingardena 3, 30-060, Kraków, Poland*

## 1. Abstract

We have designed, constructed and optimized a novel cell for micro photoacoustic spectroscopy (microPAS) as part of a collaboration between the Canadian Light Source and Natural Resources Canada. The cell allows inspection of micrometric samples under a visible light microscope, together with measurement of mid-infrared absorption spectra. Our goal is to introduce a methodology enabling mid-IR investigation of micrometric samples that are inaccessible by other measurement configurations; this includes opaque samples, as well as those that are prone to scattering or possess irregular topography. We have achieved this objective, building and testing a prototype PA spectromicroscopy cell using several different optical sources. This device was operated successfully during collaborative experiments at MIRIAM beamline B22 of Diamond Light Source in July 2016.

## 2. Experiment details

The cell for microPAS studies of microsamples has been described in proposal SM13668. The cell comprises a metal block that encloses a central cavity containing the sample. CsI or KBr optical windows, which are transparent to visible and mid-IR light, allow both inspection and irradiation of the sample using a focused IR beam. The sample cavity is connected to a channel that transmits the acoustic (pressure) wave to a sensitive microphone.

Measurements with laser excitation were performed using a pulsed 635-nm diode laser (average power ~5 mW) and a cw Quantum Cascade Laser (QCL, Daylight Solutions External Cavity QCL TLS-41060), tunable from 1565 to 1745 cm$^{-1}$. The QCL radiation was chopped mechanically at frequencies of either 10 or 20 Hz. In the experiments with both laser sources, the modulated laser light was focused onto the PA cell using an Agilent Technologies infinity corrected 15X 0.62NA Schwarzschild objective. The signal from the microphone was demodulated using a Signal Recovery lock-in amplifier (Model 7270).

Fourier transform infrared (FTIR) measurements were performed using a Vertex 80v spectrometer (Bruker Optics). The internal halogen and globar sources, or the synchrotron beam from beamline B22, were used for excitation. Light was extracted from the left port of the spectrometer and focused into the cell using an Agilent 15x Schwarzschild objective. The analog signal from the microphone was amplified and communicated to the external detector signal channel (ADC) of the spectrometer. Measurements were controlled using the OPUS (Bruker Optics) software package that operates the interferometer. The samples consisted of a 60 µm glassy carbon film, polystyrene beads (~35–75 µm, TentaGel) and acetyl polystyrene beads (~90 µm, TentaGel), and were mounted on the bottom window of the PAS cell using double-sided tape.

## 3. Results

We utilised the various light sources available at beamline B22 in a series of tests designed to evaluate the performance of the microPAS cell. We succeeded in observing and recording PA responses from the cell using the diode laser, QCL, the internal FTIR globar and halogen sources, and the synchrotron light source. Several optical configurations for focusing the exciting beam on the sample were tested. The key results are summarised in the following paragraphs.

*Diode laser and QCL measurements*

The capability of the cell to produce a microPAS response was initially tested using the 635-nm diode laser and the tunable QCL laser as excitation sources. In the first test, the 60 µm thick glassy carbon sample was irradiated using the diode laser. The microPAS signal produced by the microphone was monitored as a function of the laser pulse frequency. A plot of the signal S versus the inverse of the modulation frequency *f* (Figure 1A) shows an approximately linear relationship, which is to be expected under conditions where the sample is optically opaque and thermally thin [1]. This result confirms that the microPAS signal originates from the sample and that the cell is fully functional.

The second test investigated the capability of the cell for measurement of a microPAS infrared spectrum. We used a layer of polystyrene and acetyl polystyrene beads as the sample, manually selecting wavenumber settings available with the tunable QCL source. The microphone signal was recorded at wavenumber intervals of 5 or 10 $cm^{-1}$. Glassy carbon was used as to generate a reference (background) spectrum; consistent with the result in Figure 1A, this sample yields PA signals that depend only on incident intensity. The resulting microPAS spectra of the beads (Figure 1B, discrete points) agree well with PA FTIR spectra (continuous lines) recorded using a commercial (macro) PA cell in separate laboratory-based experiments on the same sample materials. Our observations confirm that the PA spectromicroscopy cell can be used for infrared absorption spectroscopy measurements.

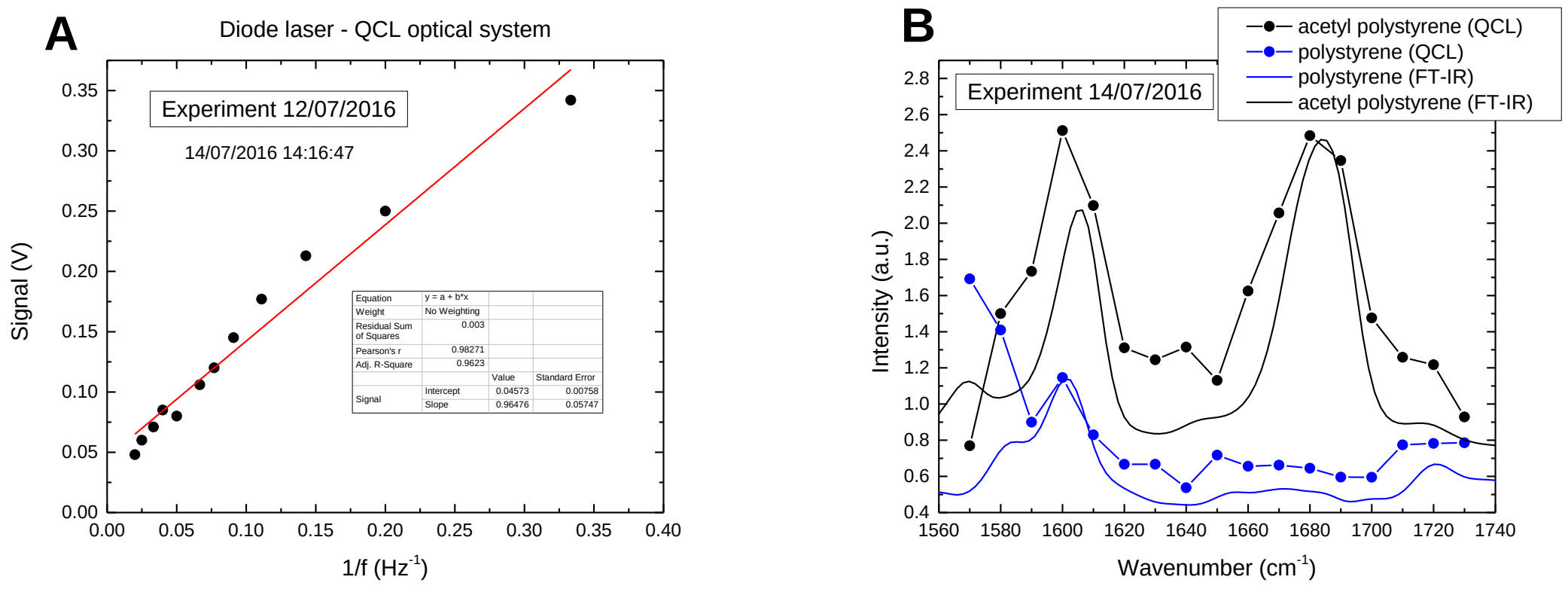

*Figure 1. A – Microphone signal S as a function of the inverse of the pulse frequency f with 635 nm excitation for a sample of glassy carbon. The linear relationship between S and 1/f indicates that the signal originates from the PA effect. B – Intensity of the reference-corrected PA spectromicroscopy signal as a function of laser excitation wavenumber (dotted lines) compared with PA FTIR spectra (continuous lines) acquired with a commercial cell for samples of polystyrene (blue) and acetyl polystyrene (black). Generally good agreement between the two sets of curves shows the functionality of the newly constructed cell.*

*FTIR measurements*

We next tested the viability of PA spectromicroscopy measurements using the Vertex 80v FTIR spectrometer on B22. The scanning of the movable interferometer mirror during conventional operation of the spectrometer provided the modulation required for generation of the PA signal. As mentioned previously, the internal halogen (NIR) and globar (MIR) sources and the external synchrotron radiation source were used to provide excitation. A 15x Schwarzschild objective focused the beam into the cell. Figure 2 shows the spectra obtained for a layer of polystyrene beads using the three sources. The intensity scale has been matched for all spectra, showing that the intensity of the signal is stronger when SR is used. The optical path used for the 3 sources was the same, but the measurements settings were not identical for the spectra acquisition. Thus in Fig.2 the evident highest spectral quality of SR should be referred to a better signal-to-noise rather than higher signal intensity.

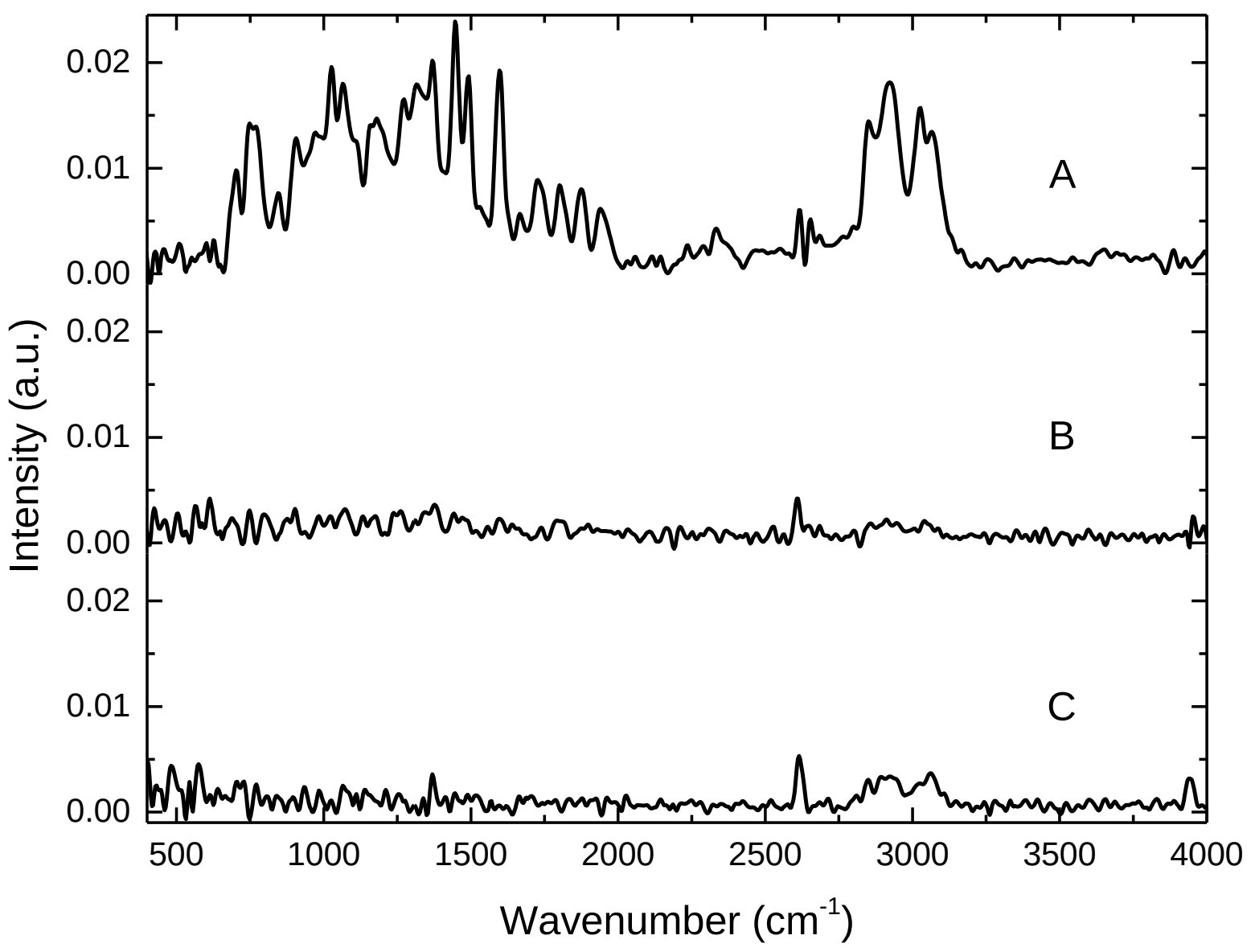

*Figure 2. microPAS FTIR spectra of a layer of polystyrene beads obtained using different excitation sources; A – Synchrotron Radiation from B22; B – Globar Lamp; C – Halogen Lamp.*

*Single Bead SR measurements*

We proved the viability of selecting single sample locations by use of a visible light portable microscope to locate and probe particular beads in a mixed layer of polystyrene and acetyl polystyrene. Centering of the probe beam on a specific location is confirmed by using the visible light component of the focused synchrotron beam. Figure 3 shows two such measurements on different beads. The appearance of a band due to carbon-oxygen functional groups confirms that bead one consists of acetyl polystyrene while the absence of this band indicates that bead 2 is polystyrene. These results accord with the spectra plotted in Figure 2B.

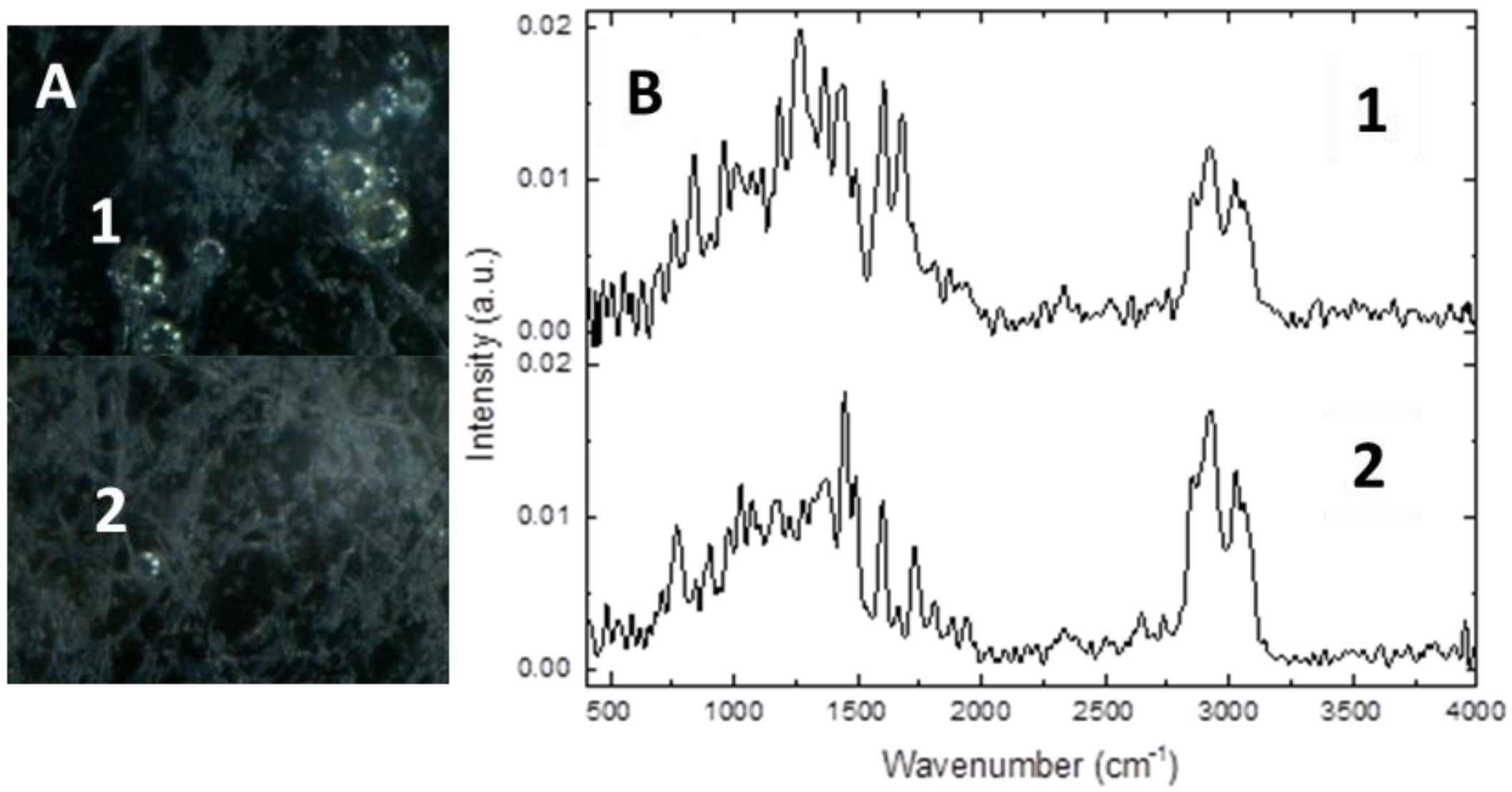

*Figure 3. A – Visible light images of two regions in a sample consisting of a mixture of polystyrene and acetyl polystyrene beads. B – micro PAS FTIR spectra of the beads numbered 1 and 2 in panel A acquired with the PA spectromicroscopy cell. Comparison of the FTIR spectra confirms that bead 1 is acetyl polystyrene and bead 2 is polystyrene.*

## 4. Conclusions and future work

We have shown that our microPAS cell enables measurement of mid-infrared spectra from micrometric samples using a variety of light sources and optical configurations. To the best of our knowledge, our work represents the first successful PA spectroscopy measurements ever performed in a microscopy configuration and constitute a proof-of-concept of feasibility of microPAS method. Future work will involve optimization of measurement conditions for SR experiments and the application to specific problems involving samples that are difficult to study via optical mid-IR microscopy configurations, such as high scattering samples e.g. microcrystals, composite microsamples/beads/fibers, and thick sample sections not suitable for microATR e.g. soft tissues or valuable sample fragments. The possibility of performing depth profiling by changing the modulation frequency of the incident light will also be explored. [1] Finally, the advantage of using a synchrotron light source over a benchtop one will be assessed.

Acknowledgments: Research described in this article was carried out with the support of Diamond Light Source (proposal SM13668-1) and of the Canadian Light Source. The Canadian Light Source is supported by the Canada Foundation for Innovation, Natural Sciences and Engineering Research Council of Canada, the University of Saskatchewan, the Government of Saskatchewan, the National Research Council Canada, and the Canadian Institutes of Health Research

Luca Quaroni is supported by the European Union’s Horizon 2020 research and innovation program under the Marie Skłodowska-Curie grant agreement No. 665778 (POLONEZ fellowship managed by the National Science Center, Poland, registration number UMO-2016/21/P/ST4/01321).